\documentclass[aps,twocolumn,pre,floatfix]{revtex4-2}

\usepackage{booktabs}

\usepackage{xcolor,hyperref}
\hypersetup{
   colorlinks,
   linkcolor={blue!50!black},%{red!80!black},
   citecolor={blue!50!black},
   urlcolor={blue!80!black}
}
\usepackage{epsfig, amsmath}
\usepackage{bm}
\usepackage{color}
\usepackage{soul,ulem}
\usepackage{hyperref}
\usepackage{footmisc}
\usepackage{wrapfig2,graphicx,enumitem}
\DeclareMathAlphabet{\mathitbf}{OML}{cmm}{b}{it}

\newcommand{\kv}{\mathitbf k}
\newcommand{\uv}{\mathitbf u}
\newcommand{\rv}{\mathitbf r}
\newcommand{\wv}{\mathitbf w}

\newcommand{\nv}{\mathitbf n}

\newcommand{\psiv}{\bm{\psi}}

\newcommand{\calBold}[1]{\mbox{\boldmath${\cal #1}$}}
\newcommand{\dbar}{{\,\mathchar'26\mkern-12mu d}}

\begin{document}

% \title{Effect of the smoothness of interatomic potential cutoffs\\ on the non-phononic VDoS of computer glasses}
\title{A method for measuring the dispersion of elastic waves in disordered computer-solids}
\author{Edan Lerner}
\email{e.lerner@uva.nl}

\affiliation{Institute for Theoretical Physics, University of Amsterdam, Science Park 904, 1098 XH Amsterdam, the Netherlands}

\begin{abstract}
The dispersion of elastic waves in disordered solids plays a key role in determining the vibrational density of states and harmonic wave attenuation rates. As such, the availability of robust computational approaches to the precise extraction of the dispersion is of key importance. Here we present a simple method --- the imposed wave method (IWM) ---, which provides direct access to the dispersion of elastic waves in computer models of solids, without any fitting involved. We directly benchmark the method against the `ground-truth' obtained from direct diagonalization of solids' hessian matrices, to find good agreement. We discuss limitations of and finite-size effects in the method, and show that exploiting the method's finite-size scaling provides access to a fundamental quantifier of mechanical disorder that determines wave attenuation rates and spectral widths.
  
\end{abstract}

\maketitle

\section{i\lowercase{ntroduction}}
\vspace{-0.2cm}

Elastic waves, also known as long-wavelength \emph{phonons}, emerge in any solid whose hamiltonian is translationally invariant~\cite{kittel2005introduction}. In the framework of continuum linear elasticity~\cite{landau_lifshitz_elasticity}, the frequency of waves of wavelength $\lambda$ is proportional to $c/\lambda$, with $c$ a wave-speed determined by the solid's elastic moduli and mass density~\cite{kittel2005introduction}. This proportionality is commonly expressed as a linear relation between the frequency of a wave and its wavenumber $k\!=\!2\pi/\lambda$ as $\omega(k)\! =\! ck$, widely known as the linear dispersion relation of elastic waves.

At lower wavelengths or higher wavenumbers, the linear dispersion of elastic waves gradually breaks down; generically (though exceptions exist, see, e.g.~\cite{PhysRevB.94.144205}), the frequency of waves with smaller wavelengths is \emph{lower} than the linear dispersion relation predicts, i.e.~waves' frequencies soften compared with the linear dispersion expectation, with decreasing wavelength. While the mechanism leading to nonlinear wave dispersion in crystals is well-understood~\cite{kittel2005introduction}, the situation in disordered solids is less clear~\cite{Schirmacher_prl_2007,monaco2009anomalous,ganter2010rayleigh,Monaco_prl_2011,chumakov2016relation,nonlinear_dispersion_arxiv_2026}. As the dispersion of elastic waves determines the form of the vibrational density of states (VDoS)~\cite{Schirmacher_prl_2007,monaco2009anomalous,ganter2010rayleigh,Monaco_prl_2011,chumakov2016relation,nonlinear_dispersion_arxiv_2026} --- which, in turn, governs the solid's mechanical and thermodynamic linear response functions --- the ability to measure wave dispersion accurately and efficiently in computer simulations is of key importance.

There are two common approaches to measuring the dispersion of elastic waves in disordered computer solids. The first involves calculating the dynamic structure factor $S(k,\omega)$ --- which features a peak at the dispersion frequency $\omega(k)$ --- and fitting it to some model (typically a Lorentzian~\cite{tanaka_boson_peak_2008,monaco2009anomalous,Schirmacher_2013_boson_peak}, but not always~\cite{tanaka_2d_modes_2022}) in order to extract the wave dispersion. Another approach involves tracking correlation functions of vibrational dynamics following an initial wave-like perturbation, and extracting the attenuation rate and frequency of waves of some given wavelength via fits to damped oscillations~\cite{lemaitre_tanaka_2016,Ikeda_scattering_2018,wang2019sound,scattering_jcp,jcp_letter_scattering_2021,grzegorz_2025_scattering_perspective}. 

Here we put forward a third approach --- the imposed wave method (IWM) --- that involves imposing a wave-like force with some wavelength on the glass, obtaining the linear response to that wave-like force, and assessing the effective frequency of the generated response. Similar procedures were put forward in~\cite{liu_soft_matter_2013,brian_prl_2017}. The method is straightforward to implement, and does not rely on any fitting to obtain the wave dispersion $\omega(k)$. We benchmark the method against the ground-truth dispersion obtain from direct diagonalization of the hessian matrix of two model disordered solids in three dimensions. Finally, we exploit finite-size effects in the IWM framework to extract the spectral widths of elastic waves~\cite{ganter2010rayleigh,nonlinear_dispersion_arxiv_2026}.

This paper is structured as follows. In Sect.~\ref{sec:models} we provide details about the model disordered solids employed for this work. Sect.~\ref{sec:method} describes in details the IWM for measuring wave dispersion in disordered solids. Sections~\ref{sec:benchmark_networks} and~\ref{sec:benchmark_glass} benchmark the IWM against calculations in disordered spring networks and in a model glass former, respectively. Sect.~\ref{sec:finite_size_effects} reviews the finite-size effects seen within the IWM framework, while Sect.~\ref{sec:widths} explains how those can be exploited to obtain spectral widths. Sect.~\ref{sec:previous_work} discusses relations to previous work, and a brief summary is provided in Sect.~\ref{sec:summary}.

\section{C\lowercase{omputer models of disordered solids}}\label{sec:models}
\vspace{-0.2cm}
We demonstrate and benchmark the IWM on two models of disordered solids, described below.

%%%%%%%%%%%%%%%%%%%%%%%%%%%%%%
%%%%%%%%%%%%%%%%%%%%%%%%%%%%%%%%%%%%%%%%%%%%%%%%%%%%%%%%%%%%%%%%%%%%%%%
\begin{figure*}[ht!]
  \includegraphics[width = 0.85\textwidth]{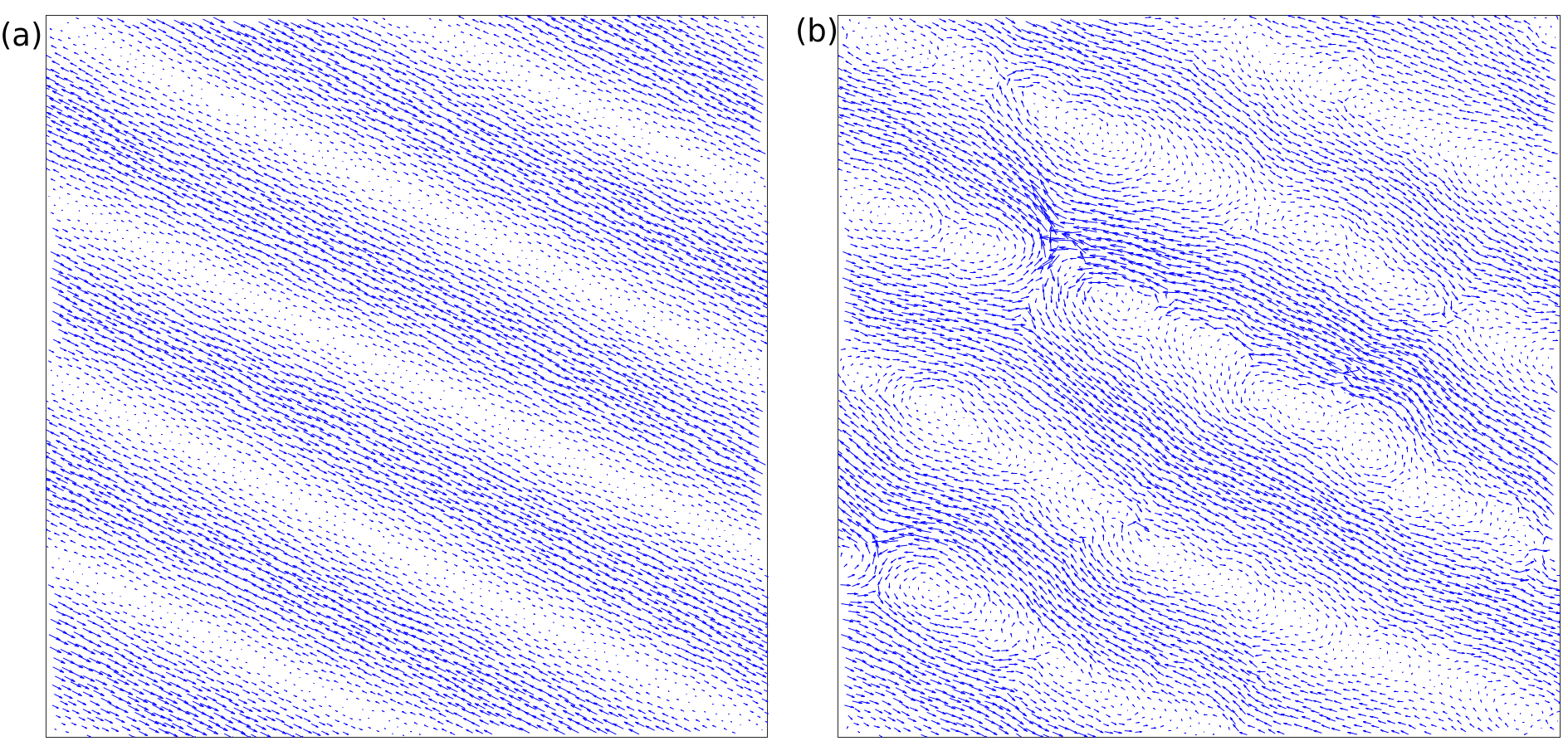}
%  \vspace{-0.5cm}
  \caption{\footnotesize  (a) An example of the imposed wave $\wv(\kv)$ of Eq.~\eqref{eq:imposed_wave} in a 2D glass of $N\!=\!16,000$ particles. (b) The linear response $\uv(\kv)$ to the imposed wave, see Eq.~\eqref{eq:linear_response}. It is a wave-like object that respects the system's structural and mechanical disorder.}
  \label{fig:fake_phonon_example}
\end{figure*}
%%%%%%%%%%%%%%%%%%%%%%%%%%%%%%%%%%%%%%%%%%%%%%%%%%%%%%%%%%%%%%%%%%%%%%

\subsection{Disordered networks of relaxed Hookean springs}
\vspace{-0.2cm}
We construct three dimensional (3D) disordered networks of Hookean springs of stiffness $\kappa$ by diluting the interaction network of a soft-sphere glass, employing an algorithm that maintains low node-to-node fluctuations of connectivity. Details about the model and algorithm can be found in~\cite{anomalous_elasticity_soft_matter_2023}. We study system sizes ranging from $N\!=\!256$ to $N\!=\!256$K nodes of equal mass $m$. For benchmarking the IWM we employ networks with a mean coordination of  $z\!=\!16.24$, which is much larger than the Maxwell threshold $z_{\rm c}\!=\!6$ in 3D~\cite{Maxwell01041864}. We also study networks with coordinations varying between $z\!=\!6.64$ to $z\!=\!16.24$. By construction of our networks, each spring resides exactly at its respective rest-length, such that the potential energy $U$ of the system vanishes identically. For this system, frequencies are reported in terms of $\sqrt{\kappa/m}$, and lengths are reported in terms of $a_0\!\equiv\!(V/N)^{1/3}$ with $V$ denoting the system's volume.

%%%%%%%%%%%%%%%%%%%%%%%%%%%%%%
%%%%%%%%%%%%%%%%%%%%%%%%%%%%%%%%%%%%%%%%%%%%%%%%%%%%%%%%%%%%%%%%%%%%%%%
\begin{figure*}[ht!]
  \includegraphics[width = 1.0\textwidth]{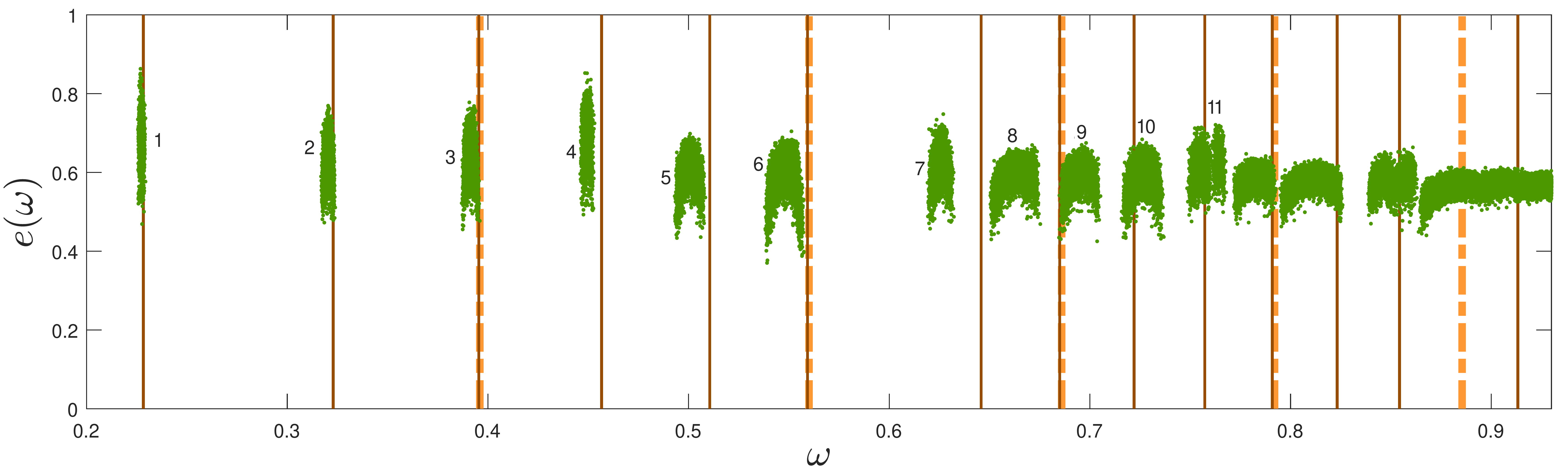}
%  \vspace{-0.5cm}
  \caption{\footnotesize  Participation ratio $e(\omega)$ (see text for definition) of vibrational modes of disordered spring networks of $N\!=\!16,000$ nodes with mean coordination $z\!=\!16.24$, scatter-plotted against those modes' frequencies $\omega$. The vertical lines mark the expected phononic frequencies based on a linear dispersion: continuous thin lines represent shear waves, and dashed thick lines represent sound waves. The softening of phononic frequencies relative to a linear dispersion is evident by the increasing discrepancy between phonon-band frequencies and the vertical lines, as higher frequencies are considered. At frequencies $\omega\!>\!\omega_\dagger(N)\!\approx\!0.9$ the gaps between subsequent bands close, see text for discussion.}
  \label{fig:participation_vs_freq_networks}
\end{figure*}
%%%%%%%%%%%%%%%%%%%%%%%%%%%%%%%%%%%%%%%%%%%%%%%%%%%%%%%%%%%%%%%%%%%%%%

\subsection{Polydisperse glass}
\vspace{-0.2cm}
We study a glass-forming model of purely repulsive soft-spheres of equal mass $m$ interacting via a $\sim\!\varepsilon(\sigma/r)^{10}$ pairwise potential, where $r$ is the pairwise distance between particles, and $\varepsilon,\sigma$ are energy and length scales, respectively. The size distribution of the spheres is tailored to optimize the efficiency of the swap-Monte-Carlo algorithm~\cite{LB_swap_prx} that allows for very deep supercooling. Details about the model can be found in~\cite{boring_paper}. For this system, frequencies are reported in terms of $\sqrt{\varepsilon/m\sigma^2}$ and lengths in terms of $\sigma$. We study systems of $N\!=\!16$K, 64K and 256K. Glasses were created by a potential-energy minimization of supercooled states equilibrated at the parent temperature $T_{\rm p}\!=\!0.40\varepsilon/k_{\rm B}$. The crossover temperature for this glass-former is estimated at $0.75\varepsilon/k_{\rm B}$~\cite{cge2_jcp2020}, implying that our glasses are formed from moderately supercooled liquid states.

\section{T\lowercase{he `imposed wave' method} (IWM)}\label{sec:method}
\vspace{-0.2cm}
Disordered solids' structural and mechanical disorder dresses elastic waves (phonons) with noise~\cite{phonon_widths}. As a result, the degeneracy that waves of the same wavelength would feature in a homogeneous, isotropic continuum -- is lifted~\cite{phonon_widths}. In order to mimic the disorder-induced noise in waves, we adopt the following approach.

Given some wavevector $\kv$, we first build a wave-like force $\wv(\kv)$ as
\begin{equation}
    \wv_i^{({\rm s},\ell)}(\kv) = \hat{\kv}^{({\rm s},\ell)}\,e^{i\kv\cdot\rv_i}-N^{-1}\sum_j\hat{\kv}^{({\rm s},\ell)}\,e^{i\kv\cdot\rv_j}\,,
\label{eq:imposed_wave}
\end{equation}
where the index $i$ labels the $i^{\mbox{\tiny th}}$ particle (in glasses) or node (in networks), $\rv_i$ denotes the coordinates of the $i^{\mbox{\tiny th}}$ particle or node,  $\kv$ is an admissible wavevector and $\hat{\kv}^{({\rm s},\ell)}$ are polarization unit vectors corresponding to shear and longitudinal waves, respectively. The polarization of longitudinal phonons reads $\hat{\kv}^{({\ell})}\!=\!\kv/|\kv|$ and hence that of shear phonons is $\hat{\kv}^{({\rm s})}\cdot\hat{\kv}^{(\ell)}\!=\!0$ (there are $\dbar\!-\!1$ shear polarizations, with $\dbar$ denoting the dimension of space). The admissible wavevectors (in our system with periodic boundary conditions) are $\kv\!=\!2\pi\nv/L$, where $L$ the system's linear size, and $\nv\!=\!(n_x,n_y,n_z)$ is a vector of integers. The second term on the RHS of Eq.~(\ref{eq:imposed_wave}) ensures that $\wv(\kv)$ has no projection on the $\dbar$ center of mass translations (zero modes). For the sake of brevity, we omit hereafter the superscripts and subscripts $({\rm s},\ell)$ from all subsequent definitions, but they are implicitly implied. An example of $\wv(\kv)$ in a two-dimensional (2D) glass is shown in Fig.~\ref{fig:fake_phonon_example}a (we show a 2D image for illustrational purposes; all subsequent analyses in this work are performed in 3D models).

The linear response to the imposed wave $\wv(\kv)$ is given by
\begin{equation}
    \uv(\kv) = \calBold{H}^{-1}\cdot\wv(\kv)\,.
\label{eq:linear_response}
\end{equation}
where $\calBold{H}^{-1}$ is the inverse of the Hessian matrix $\calBold{H}_{ij}\!\equiv\!\frac{\partial^2U}{\partial\rv_i\partial\rv_j}$. An example of $\uv(\kv)$ is shown in Fig.~\ref{fig:fake_phonon_example}b, demonstrating that its spatial structure closely resembles a phononic vibrational mode -- it is a wave dressed by noise. If $\uv(\kv)$ were an exact eigenfunction of $\calBold{H}$, its eigenvalue would be the associated frequency squared $\omega^2$. In complete analogy, we associate with $\uv(\kv)$ the frequency squared $[\omega(\kv)]^2$ according to
\begin{equation}
    [\omega(\kv)]^2 = \frac{\uv(\kv)\cdot\calBold{H}\cdot\uv(\kv)}{\uv(\kv)\cdot\uv(\kv)} = \frac{\wv(\kv)\cdot\calBold{H}^{-1}\cdot\wv(\kv)}{\wv(\kv)\cdot\calBold{H}^{-2}\cdot\wv(\kv)}\,.
\label{eq:wave_freq}
\end{equation}
The dispersion $\omega(k)$ is obtained by averaging $\omega(\kv)$ of Eq.~\eqref{eq:wave_freq} over various wavevectors $\kv$ that share the same magnitude $k\!=\!|\kv|$ and also over independent disordered solid realizations. 

\section{S\lowercase{pring networks benchmark of the} IWM}\label{sec:benchmark_networks}
\vspace{-0.2cm}
Testing the IWM in the cleanest way requires analyzing a disordered solid whose low-frequency vibrations are exclusively phonons, i.e.~free of non-phononic modes. A unique example of such a system is a highly-coordinated disordered spring network in which each spring resides exactly at its rest length, such that the potential energy of the solid vanishes identically. This model solid is frustration free, which results in a gapped nonphononic spectrum, with an onset frequency $\omega_\star\!\sim\!z\!-\!z_{\rm c}$~\cite{mw_EM_epl}, where $z$ is the average coordination of the network nodes, and $z_{\rm c}\!=\!2\dbar$ is the Maxwell threshold~\cite{Maxwell01041864}, with $\dbar$ denoting the dimension of space. Below the onset frequency $\omega_\star$ of nonphononic modes, one expects to only find phonons, which makes this system a perfect testbed for the IWM.

In Fig.~\ref{fig:participation_vs_freq_networks} we scatter plot the participation ratio $e\!\equiv\!\big[N\sum_i(\psiv_i\cdot\psiv_i)^2\big]^{-1}$ of vibrational modes $\psiv$ vs.~those modes' vibrational frequencies $\omega$, for disordered spring networks of $N\!=\!16,000$ nodes. We indeed see, as expected, that low-frequency modes are grouped into discrete phonon-bands, with sizeable gaps between the bands, up until a crossover frequency $\omega_\dagger\!\sim\!N^{-2/15}$~\cite{phonon_widths} (in 3D; $\omega_\dagger\!\sim\!N^{-2/(\dbar(\dbar+2))}$ for solids in $\dbar$ dimensions) above which the phonon-band widths surpass the gaps between the bands ($\omega_\dagger\!\approx\!0.9$ for our networks of $N\!=\!16,000$, cf.~Fig.~\ref{fig:participation_vs_freq_networks}). The discretization of phonon bands at low frequencies allows one to identify the wavenumber $k$ associated with each band, and to obtain the average frequency $\omega(k)$ per band, see Appendix~\ref{sec:appendix} for details. Some bands (see, e.g., bands 3,6,8, and 11 in Fig.~\ref{fig:participation_vs_freq_networks}) involve combinations of shear and sound waves; we omit those bands from our analysis since we cannot disentangle the associated hybridized modes. 

%%%%%%%%%%%%%%%%%%%%%%%%%%%%%%%%%%%%%%%%%%%%%%%%%%%%%%%%%%%%%%%%%%%%%%%
\begin{figure}[ht!]
  \includegraphics[width = 0.5\textwidth]{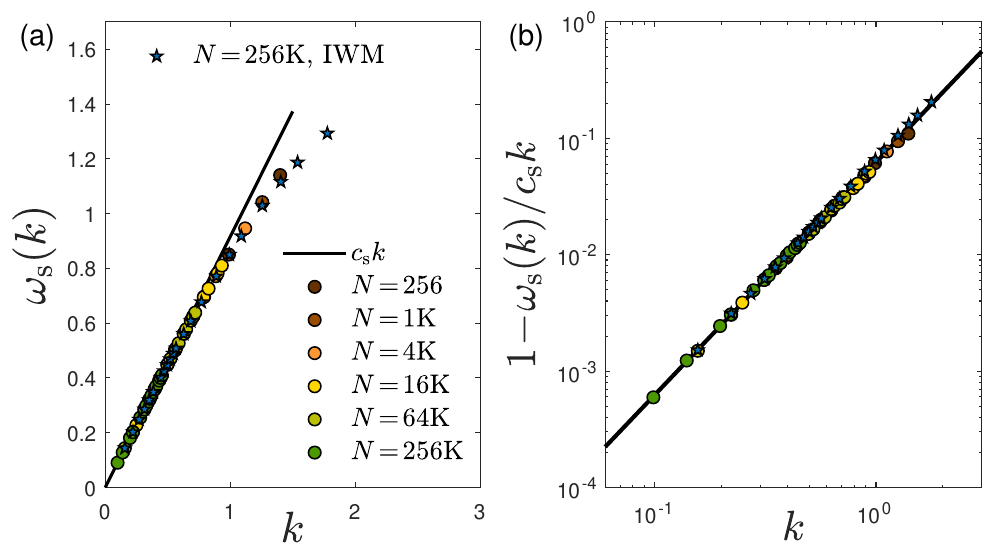}
%  \vspace{-0.5cm}
  \caption{\footnotesize  (a) The `ground truth' shear-wave dispersion $\omega_{\rm s}(k)$ measured from vibrational modes' frequencies (circles), superimposed with the result of the IWM (stars). (b) The difference $1\!-\!\omega_{\rm s}(k)/c_{\rm s}k$ is plotted against $k$ for the same datasets of panel (a), establishing that the first nonlinear correction to the linear dispersion is cubic (as required), that the prefactor of the cubic correction is $N$-independent, and that the IWM agrees well with the ground-truth.}
  \label{fig:networks_benchmark}
\end{figure}
%%%%%%%%%%%%%%%%%%%%%%%%%%%%%%%%%%%%%%%%%%%%%%%%%%%%%%%%%%%%%%%%%%%%%%

The averages $\omega(k)$ constitute the `ground-truth' of the elastic wave dispersion, and are plotted in Fig.~\ref{fig:networks_benchmark} for various $N$ as indicated by the figure legend. We find that, at the frequencies probed, the ground-truth dispersion appears to be $N$-independent. We next superimpose the results of the IWM on the ground truth in Fig.~\ref{fig:networks_benchmark}a, to find excellent agreement. Further support for the accuracy of the IWM is seen in Fig.~\ref{fig:networks_benchmark}b in which the difference $1\!-\!\omega_{\rm s}(k)/c_{\rm s}k$ is plotted vs.~$k$, where $c_{\rm s}$ is the shear wave speed  . Since the dispersion must be an odd function of $k$~\cite{kittel2005introduction}, this difference must scale as $k^2$, in agreement with our findings. Our data further indicate that the cubic correction to the dispersion is $N$-independent, and agrees between the ground truth and the IWM results.

\section{G\lowercase{lass benchmark of the} IWM}\label{sec:benchmark_glass}
\vspace{-0.2cm}

We next turn to testing the IWM in a structural glass model. Glasses formed by quenching a liquid generically feature nonphononic vibrational modes that follow a gapless nonphononic VDoS ${\cal D}_{\rm G}(\omega)\!\sim\!\omega^4$~\cite{JCP_Perspective}. Consequently, it is not possible to cleanly identify phonon bands as was possible for the frustration-free disordered networks studied in the previous Section. We therefore adopt a different approach, described below.

%%%%%%%%%%%%%%%%%%%%%%%%%%%%%%%%%%%%%%%%%%%%%%%%%%%%%%%%%%%%%%%%%%%%%%%
\begin{figure}[ht!]
  \includegraphics[width = 0.5\textwidth]{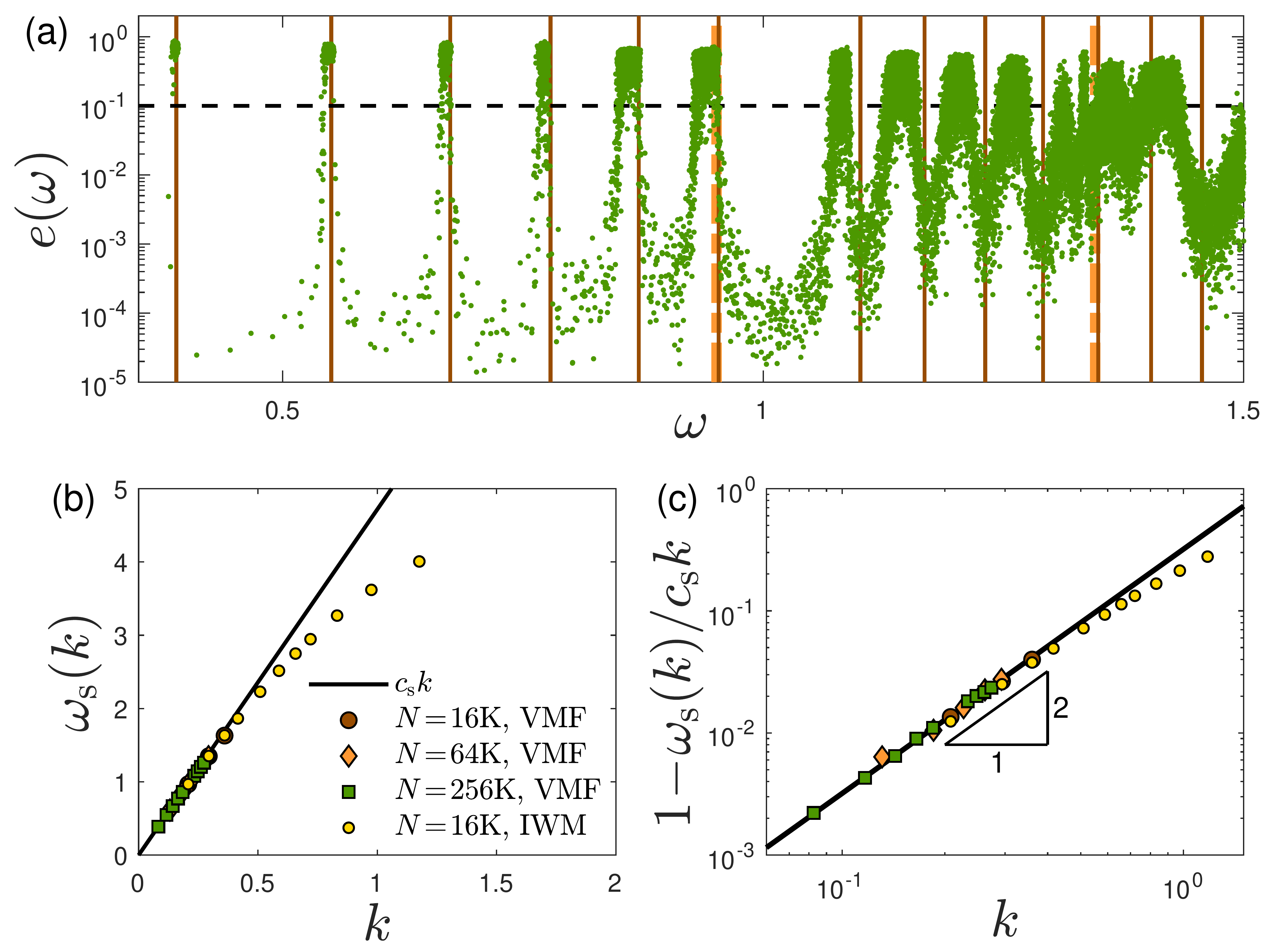}
%  \vspace{-0.5cm}
  \caption{\footnotesize  (a) The participation ratio $e(\omega)$ scatter-plotted against frequency $\omega$ of low-frequency vibrational modes in the polydisperse glass model with $T_{\rm p}\!=\!0.40$ and $N\!=\!256,000$ particles. The continuous vertical lines represent the estimations of phonon band frequencies assuming a linear dispersion. The dashed vertical lines represent sound waves. The horizontal dashed line marks the chosen threshold for identifying phonons in each band, see text for additional details. (b) The `ground truth' shear-wave dispersion $\omega_{\rm s}(k)$ measured from vibrational modes' frequencies (VMF in the figure legend), superimposed with the result of the IWM. (c) The difference $1\!-\!\omega_{\rm s}(k)/c_{\rm s}k$ is plotted against $k$ for the same datasets of panel (b), establishing that the first nonlinear correction to the linear dispersion is cubic (as required), that the prefactor of the cubic correction is $N$-independent, and that the IWM agrees well with the ground truth.}
  \label{fig:glass_benchmark}
\end{figure}
%%%%%%%%%%%%%%%%%%%%%%%%%%%%%%%%%%%%%%%%%%%%%%%%%%%%%%%%%%%%%%%%%%%%%%

In Fig.~\ref{fig:glass_benchmark}a we scatter-plot the participation ratio $e(\omega)$ of vibrational modes $\psiv$ vs.~those modes' vibrational frequencies $\omega$, for glasses of $N\!=\!256,000$ particles. The quantization of low-frequency phonon-bands is apparent; however, quasilocalized modes with $e\!\sim\!1/N$ appear in between phonon-bands~\cite{phonon_widths,JCP_Perspective}. In order to obtain the ground-truth wave dispersion, we introduce a threshold participation ratio   $e_{\rm th}\!=\!0.1$ --- marked by the horizontal dashed line in Fig.~\ref{fig:glass_benchmark}a---, and take the average frequency of modes with $e(\omega)\!>\!e_{\rm th}$ within each discrete phonon band, see Appendix~\ref{sec:appendix} for further details. Also in this case, we omit mixed sound/shear bands (see, e.g.,~the 6th band in Fig.~\ref{fig:glass_benchmark}a) from our analysis, and consider bands with frequencies $\omega\!<\!\omega_\dagger(N)$ ($\omega_\dagger\!\approx\!1.3$ for the $N\!=\!256,000$ data of Fig.~\ref{fig:glass_benchmark}a).

We applied the analysis described above to systems of $N\!=\!16$K, 64K and 256K; the results of this analysis are presented in Fig.~\ref{fig:glass_benchmark}b, where we plot the dispersion of shear waves $\omega_{\rm s}(k)$ vs.~wavenumber $k$. We superimpose the results of the IWM applied to glasses of $N\!=\!16$K particles, to find very good agreement. Plotting $1\!-\!\omega_{\rm s}(k)/c_{\rm s}k$ in Fig.~\ref{fig:glass_benchmark}c reveals --- as in the case of our disordered networks --- that the cubic correction to the dispersion is $N$-independent, and appears to be the same for the dispersion obtained via the ground truth and via the IWM.

\section{F\lowercase{inite-size effects}}\label{sec:finite_size_effects}
\begin{figure}[ht!]
  \includegraphics[width = 0.5\textwidth]{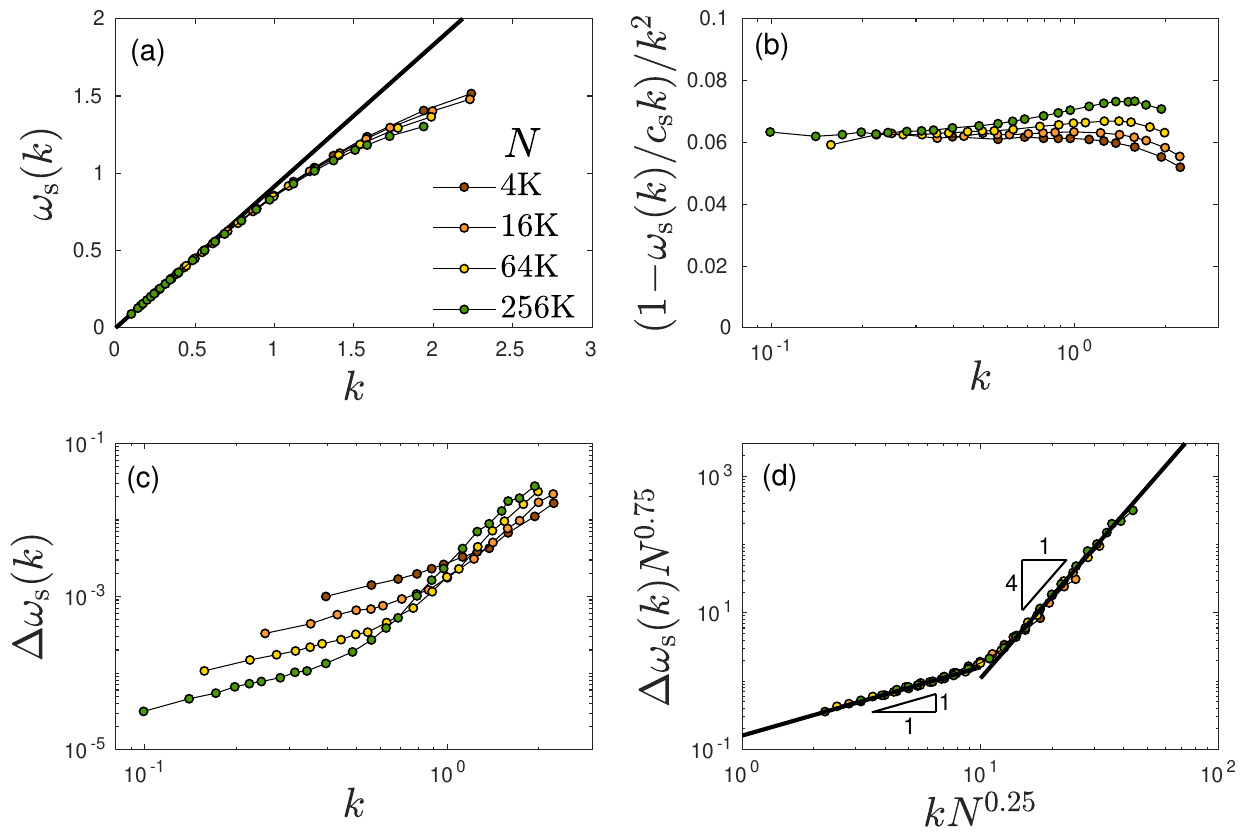}
%  \vspace{-0.5cm}
  \caption{\footnotesize (a) The shear-wave dispersion $\omega_{\rm s}(k)$ calculated via the IWM shows clear $N$-dependence at large $k$. However, the rescaled differences $(1\!-\!\omega_{\rm s}(k)/c_{\rm s}k)/k^2\!=\!{\cal B}_{\rm s}\!+\!{\cal O}(k^2)$ --- plotted in panel (b) --- reveal that the coefficient ${\cal B}_{\rm s}$ of the cubic correction $-c_{\rm s}{\cal B}_{\rm s}k^3$ to the linear dispersion is $N$-independent. (c) The widths $\Delta\omega_{\rm s}(k)$ of imposed-wave frequencies are plotted against the wavenumber $k$. (d) A scaling collapse of the widths $\Delta\omega_{\rm s}(k)$ reveal a crossover at a wavenumber $k_\star\!\sim\!N^{-0.25}$ to a $\sim\!k^4$ Rayleigh-like scaling, see text for discussion.}  
  \label{fig:finite-size_fig}
\end{figure}

\vspace{-0.2cm}
We next study finite-size effects in the IWM framework. Fig.~\ref{fig:finite-size_fig}a plots the dispersion $\omega_{\rm s}(k)$ of shear waves measured via the IWM framework in our random spring networks with coordination $z\!=\!16.24$ and various system sizes as indicated by the figure legend. We see that, at wavenumbers $k\!>\!1$ the dispersion shows a clear $N$-dependence, with larger systems featuring a sharper softening as $k$ increases. These data indicate that the IWM is not reliable at high wavenumbers $k$.

%%%%%%%%%%%%%%%%%%%%%%%%%%%%%%%%%%%%%%%%%%%%%%%%%%%%%%%%%%%%%%%%%%%%%%%
%%%%%%%%%%%%%%%%%%%%%%%%%%%%%%%%%%%%%%%%%%%%%%%%%%%%%%%%%%%%%%%%%%%%%%

That said, in Fig.~\ref{fig:finite-size_fig}b we plot the rescaled differences $(1\!-\!\omega_{\rm s}(k)/c_{\rm s}k)/k^2\!=\!{\cal B}_{\rm s} + {\cal O}(k^2)$ vs.~wavenumber $k$. We find that the signals all converge to a constant at low-$k$, indicating that the cubic coefficient ${\cal B}_{\rm s}$ --- which has dimensions of length$^2$~\cite{nonlinear_dispersion_arxiv_2026} --- appears to be $N$-independent, at least within the accuracy of our data. We conclude that the IWM can reliably  provide the cubic correction to the linear dispersion of elastic waves, but not higher-order terms.

Recall next that the IWM associates a frequency $\omega_{\rm s}(\kv)$ to each trial wave $\uv_{\rm s}(\kv)$ generated (cf.~Eqs.~(\ref{eq:linear_response}),(\ref{eq:wave_freq})), and the dispersion $\omega_{\rm s}(k)$ is obtained by averaging over all frequencies $\omega_{\rm s}(\kv)$ of all generated trial waves $\uv_{\rm s}(\kv)$. In Fig.~\ref{fig:finite-size_fig}c we plot the standard deviations $\Delta\omega_{\rm s}(k)$ (referred to below simply as `widths') of the measured set of frequencies vs.~wavenumber $k$ for the same networks as analyzed in panels (a),(b). The data indicates an $N$-dependent crossover in the scaling with $k$ of the widths. Plotting $\Delta\omega_{\rm s}(k)N^{0.75}$ vs.~$kN^{0.25}$ in Fig.~\ref{fig:finite-size_fig}d leads to a data collapse, indicating that the crossover wavenumber $k_\star\!\sim\!N^{-0.25}$. Below the crossover, the widths scale linearly with $k$, whereas above it they feature a Rayleigh-like scaling $\sim\!k^4$. Furthermore, the naive expectation $\Delta\omega_{\rm s}\!\sim\!1/\sqrt{N}$ appears to be satisfied in the small $k\!<\!k_\star$ regime, setting a relation between the exponents that result in a scaling collapse, namely $0.75\!-\!0.25\!=\!1/2$.

We speculate that the crossover in the $k$-scaling of the widths is related to the crossover seen in scattering rates from a finite-size regime to an $N$-independent regime at $k_\dagger\!\sim\!\omega_\dagger/c_{\rm s}\!\sim\!N^{-2/15}$ in 3D~\cite{phonon_widths,scattering_jcp}. However the scaling exponent we find is larger by nearly a factor of 2 than this naive expectation. We leave the resolution of this scaling law to future work.

\section{S\lowercase{pectral widths}}\label{sec:widths}
\vspace{-0.2cm}

Wave attenuation rates --- also referred to as spectral widths --- in 3D disordered solids have been shown to follow Rayleigh scattering as~\cite{Schirmacher_prl_2007,ganter2010rayleigh,scattering_jcp,jcp_letter_scattering_2021,grzegorz_2025_scattering_perspective}
\begin{equation}\label{eq:Rayleigh}
    \Gamma(k) \sim c\,a_0^3\,\chi^2 k^4\,,\quad \mbox{for}\ k>k_\dagger(N)\,,
\end{equation}
where $a_0$ is a typical interparticle length, and $\chi$ is a broadly applicable quantifier of mechanical disorder determined by spatial fluctuations of elastic moduli~\cite{karina_chi_paper_2023},   related to the `disorder parameter' $\gamma\!\sim\!\chi^2$ of the Heterogeneous Elasticity Theory~\cite{phonon_widths2}. 

Additionally, in Refs.~\cite{phonon_widths,phonon_widths2,scattering_jcp} it was shown that the widths $\Delta\omega^{\rm ph}$ of discrete phonon-bands of frequency $\omega$ in the finite-size regime $\omega\!<\!\omega_\dagger(N)$ (cf.~Fig.~\ref{fig:participation_vs_freq_networks}) are also related to the disorder quantifier $\chi$, as 
\begin{equation}\label{eq:phonon_band_widths}
    \Delta\omega^{\rm ph}\sim \chi\omega\sqrt{n_\omega/N}\,,
\end{equation}
% \begin{equation}\label{eq:phonon_band_widths}
%     \frac{\Delta\omega^{\rm ph}\sqrt{N}}{\omega\sqrt{n_\omega}} \sim\chi\,,
% \end{equation}
where $n_\omega$ denotes the degeneracy of the phonon-band with frequency $\omega$~\cite{phonon_widths}.

\begin{figure}[ht!]
  \includegraphics[width = 0.45\textwidth]{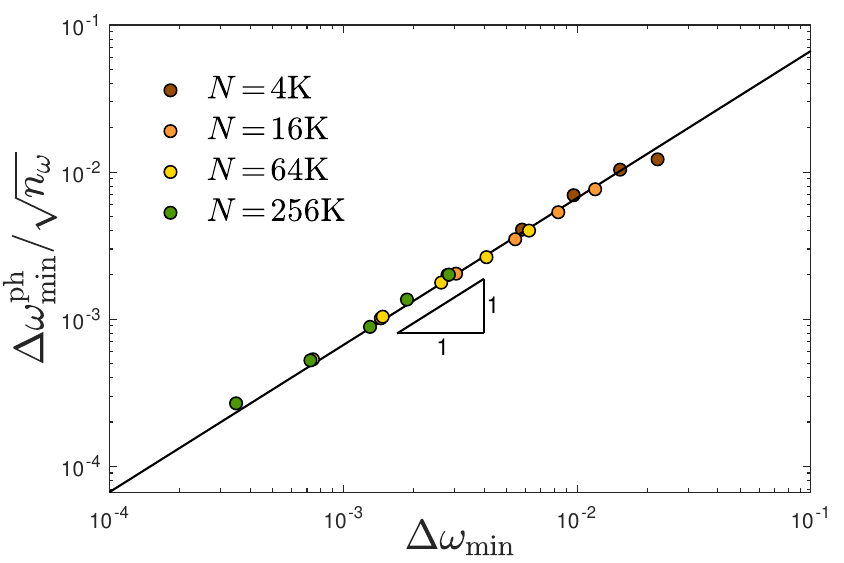}
%  \vspace{-0.5cm}
  \caption{\footnotesize Testing Eq.~(\ref{eq:IWN_widths}) that states that IWM widths $\Delta\omega$ --- similar to phonon-band widths --- are also proportional to the disorder quantifier $\chi$, see text for further details and discussion. The calculation was performed on the disordered spring networks of coordinations $z\!=\!6.64,7.28,8.56,11.12$ and $16.24$, and various system sizes as indicated by the figure legend.}  
  \label{fig:widths}
\end{figure}

Examining Eq.~(\ref{eq:phonon_band_widths}), and comparing against the scaling behavior of IWM widths $\Delta\omega(k)\!\sim\!ck/\sqrt{N}\!\sim\!\omega(k)/\sqrt{N}$ in the finite-size regime $k\!<\!k_\star$ (cf.~Fig.~\ref{fig:finite-size_fig}), one might expect that IWM widths are proportional to $\chi$ as well, namely that
\begin{equation}\label{eq:IWN_widths}
\Delta\omega(k)\sim\chi\omega(k)/\sqrt{N}\,.
\end{equation}
If true, then IWM widths $\Delta\omega(k)$ and phonon-band widths $\Delta\omega^{\rm ph}$ should be proportional to each other, up to a factor of $\sqrt{n_\omega}$. This expectation is tested in Fig.~\ref{fig:widths}, where we plot the lowest-frequency phonon-band widths $\Delta\omega^{\rm ph}_{\rm min}$ of disordered networks of coordinations $z\!=\!6.64,7.28,8.56,11.12$ and $16.24$, and for systems of various sizes as indicated by the legend --- divided by $\sqrt{n_\omega}\!=\!\sqrt{12}$ ---, vs.~the IWM widths $\Delta\omega_{\rm min}$ of shear-waves pertaining to the lowest admissible wavenumber $k_{\rm min}(N)\!<\!k_\star(N)$ for each system size considered, i.e.~all within the finite-size regime, cf.~Fig.~\ref{fig:finite-size_fig}. We find that the two widths are indeed proportional, with a proportionality coefficient of order unity, implying that Eq.~(\ref{eq:IWN_widths}) holds, and that relative variations in the disorder quantifier $\chi$ --- in addition to nonlinear wave dispersion --- can be obtained within the IWM framework as well.

%%%%%%%%%%%%%%%%%%%%%%%%%%%%%%%%%%%%%%%%%%%%%%%%%%%%%%%%%%%%%%%%%%%%%%%
%%%%%%%%%%%%%%%%%%%%%%%%%%%%%%%%%%%%%%%%%%%%%%%%%%%%%%%%%%%%%%%%%%%%%%

% Suggested placement: immediately after Sect. III (the IWM definition).
% Replace the cite keys and the equation labels \eqref{eq:imposed_wave},
% \eqref{eq:response}, \eqref{eq:omega} with those used in the manuscript
% for Eqs. (1), (2), (3).

\section{R\lowercase{elation to previous work}}
\label{sec:previous_work}

Two earlier constructions are closely related to the IWM; here we spell out how the present formulation differs from each.

%liu_soft_matter_2013,brian_prl_2017

Ref.~\cite{liu_soft_matter_2013} regards the periodic simulation cell as the
unit cell of an infinite superlattice, and diagonalizes the resulting
Bloch dynamical matrix at arbitrary wavevectors in the superlattice's
first Brillouin zone. This yields a band structure $\omega_n(\bm{k})$
whose lowest branches are transverse and longitudinal acoustic waves.
The accessible wavevectors, however, satisfy $|\bm{k}|\!<\!\pi/L$, i.e.~they
correspond to wavelengths \emph{longer} than the system, and the
procedure was devised to assess the stability of soft-sphere packings against
perturbations of their boundaries. The regime in which nonlinear
dispersion develops, $k\sim 1/a_0$, is not accessed. 

Ref.~\cite{brian_prl_2017} imposes a sinusoidal force density on a packing and computes the linear response to it, precisely as in Eq.~\eqref{eq:linear_response}. The response is then projected back onto the
imposed sinusoid, and the ratio of the projected displacement amplitude to the forcing amplitude defines a static, wavenumber-dependent
compliance that quantifies nonlocal elasticity. In our notation,
converting that compliance into a frequency amounts to
\begin{equation}
\label{eq:omega_compliance}
[\omega_{\rm C}(\bm{k})]^2 =
\frac{\bm{w}(\bm{k})\cdot\bm{w}(\bm{k})}
     {\bm{w}(\bm{k})\cdot\calBold{H}^{-1}\cdot\bm{w}(\bm{k})}\,.
\end{equation}

The IWM departs from Eq.~\eqref{eq:omega_compliance} in what is
extracted from the response. Rather than projecting $\bm{u}(\bm{k})$
back onto the imposed wave, we regard $\bm{u}(\bm{k})$ --- including
its disorder-induced dressing, cf.~Fig.~\ref{fig:fake_phonon_example}b --- as a
trial vibrational mode, and assign to it its Rayleigh quotient, cf.~Eq.~\eqref{eq:wave_freq}. The two estimators are not equivalent: applying
the Cauchy--Schwarz inequality to the vectors $\bm{w}$ and
$\calBold{H}^{-1}\!\cdot\bm{w}$ gives
$(\bm{w}\cdot\calBold{H}^{-1}\cdot\bm{w})^2 \le
 (\bm{w}\cdot\bm{w})\,(\bm{w}\cdot\calBold{H}^{-2}\cdot\bm{w})$,
and hence
\begin{equation}
\label{eq:cs_bound}
[\omega(\bm{k})]^2 \le [\omega_{\rm C}(\bm{k})]^2\,,
\end{equation}
with equality if and only if $\bm{w}(\bm{k})$ is an exact eigenvector
of $\calBold{H}$. The content of this inequality is transparent in the
eigenmode representation: denoting by $c_n\equiv\bm{\psi}_n\!\cdot\!\bm{w}$
the overlaps of the imposed wave with the vibrational modes
$\bm{\psi}_n$ of frequencies $\omega_n$ (zero modes are excluded by
construction, cf.~Eq.~\eqref{eq:imposed_wave}), one finds
$[\omega_{\rm C}]^2=\sum_n c_n^2\big/\sum_n c_n^2\omega_n^{-2}$,
a weighted harmonic mean of the $\omega_n^2$, whereas
$[\omega]^2=\sum_n c_n^2\omega_n^{-2}\big/\sum_n c_n^2\omega_n^{-4}$.
The two coincide only when the response is a pure plane wave, so their
difference is precisely the effect of the disorder-induced dressing that
the projection in Eq.~\eqref{eq:omega_compliance} discards.

\section{S\lowercase{ummary}}\label{sec:summary}
\vspace{-0.2cm}
In this work we put forward and tested the `imposed-wave method' (IWM) for measuring the dispersion of elastic waves in disordered solids. The method's advantage compared to alternative approaches is its simplicity -- it requires no fitting or modeling. The method was used extensively in Ref.~\cite{nonlinear_dispersion_arxiv_2026}, where the dispersion of 21 different ensembles of disordered solids was analyzed. We benchmarked the method against the `ground-truth' obtained via direct diagonalization of the hessian matrices of two model disordered solids, to find very good agreement at low wave numbers. At higher wave numbers, finite-size effects are seen, and the method becomes unreliable. Nevertheless, we showed that the first nonlinear correction to the linear dispersion of waves obtained with the IWM agrees with the ground-truth, and is $N$-independent. We further demonstrated how spectral widths can be obtained --- up to a multiplicative constant --- within the IWM framework, by exploiting its finite-size scaling behavior. Finally, while we only presented data for shear waves, we have checked that the behavior of longitudinal waves within the IWM is qualitatively the same (see also~\cite{nonlinear_dispersion_arxiv_2026}).

\vspace{-0.2cm}
\section*{A\lowercase{cknowledgements}}
\vspace{-0.2cm}
Discussions with Eran Bouchbinder are warmly acknowledged. This work was carried out on computational facilities generously provided by the Faculty of Chemistry of the Weizmann Institute of Science, Israel.  

\section*{D\lowercase{ata availability}}
\vspace{-0.2cm}
The data that support the findings of this article are available from
the corresponding author upon reasonable request.

\appendix

\section{`Ground truth' dispersion}\label{sec:appendix}
\vspace{-0.2cm}

In this Appendix we explain in more detail how the `ground truth' dispersion is calculated in Sects.~\ref{sec:benchmark_networks} and \ref{sec:benchmark_glass}. The key point in the following procedures is that obtaining the ground truth is only possible when low-frequency phonon bands in the VDoS of the system considered are well-separated from each other over the frequency axis, namely for frequencies $\omega\!<\!\omega_\dagger(N)$, where $\omega_\dagger(N)\!\sim\!N^{-2/15}$ (in 3D) is precisely the crossover frequency above which the width of phonon bands --- which grows with frequency --- becomes comparable to the gaps between consecutive bands~\cite{phonon_widths}.

Once phonon bands are distinguishable, one identifies the wave numbers $k$ of each band by considering the number of solutions $\#(q)$ to the sum-of-integer squares problem, namely
\begin{equation}\label{eq:sum_of_squares}
    n_x^2 + n_y^2 + n_z^2 = q\,,
\end{equation}
where $n_x,n_y,n_z$ and $q\!>\!0$ are integers~\cite{phonon_widths}. Notice in particular that Eq.~(\ref{eq:sum_of_squares}) has \emph{no} solutions for $q\!=\!7,15,23,\ldots$, i.e.~for those $q$'s, $\#(q)\!=\!0$. With this information in hand, we assign indices $q$ with nonzero $\#(q)\!>\!0$  consecutively to phonon bands in the VDoS, paying attention to the distinction between sound and shear waves. The latter is done by considering the linear dispersion represented by vertical lines in Figs.~\ref{fig:participation_vs_freq_networks} and \ref{fig:glass_benchmark}. The wavenumber is finally determined via $k\!=\!2\pi\sqrt{q}/L$ for those bands with $\#(q)\!>\!0$.

\subsection{Spring networks' ground truth dispersion}
\vspace{-0.2cm}
We consider relaxed disordered spring networks at coordination $z\!=\!16.24$, which is much higher than the Maxwell threshold $z_{\rm c}\!=\!6$ (in 3D). This means that nonphononic modes --- whose typical frequency scales as $z\!-\!z_{\rm c}$ --- are pushed to high frequencies, cleanly exposing the phonon bands, cf.~Fig.~\ref{fig:participation_vs_freq_networks}. In this scenario the only vibrational modes at low frequencies are phonons, and therefore we can obtain the ground truth dispersion by assigning a wavenumber $k$ to each band as described above, and setting $\omega(k)$ as the average frequency of the phonons in each corresponding band.

\subsection{Glasses' ground-truth dispersion}
\vspace{-0.2cm}

\begin{figure}[ht!]
  \includegraphics[width = 0.45\textwidth]{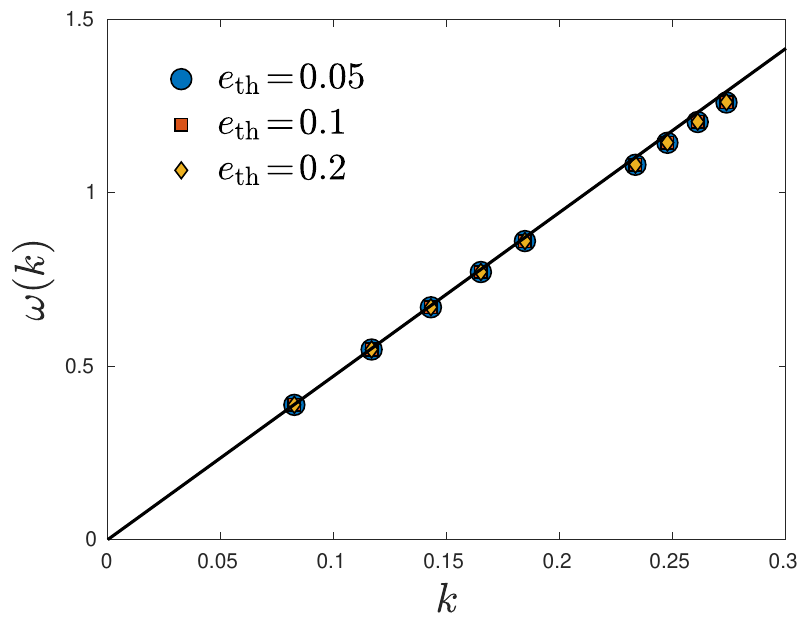}
%  \vspace{-0.5cm}
  \caption{\footnotesize The `ground-truth' dispersion obtained directly from vibrational modes of our glass model, using systems of $N\!=\!256,000$, cf.~Fig.~\ref{fig:glass_benchmark}a. These data demonstrate that the obtained dispersion is robust against changes of the choice of the threshold $e_{\rm th}$ of the participation ratio $e$, see text for discussion.}\label{fig:e_th_independence}
\end{figure}

In the glass case one needs to take care of distinguishing phononic from nonphononic vibrations when evaluating the frequency of phonon bands. We do this by selecting a threshold $e_{\rm th}$ on the participation ratio $e$ of phonons; due to their extended character, phonons typically have $e\!\sim\!{\cal O}(1)$. To obtain the ground-truth dispersion for our glasses, we assign wavenumbers $k$ to each phonon band as explained above, while again paying attention to which bands are sound vs.~shear waves via their linear dispersion. We then average the frequencies of  all the vibrational modes with $e\!>\!e_{\rm th}$ in the frequency range corresponding to each individual phonon band. In Fig.~\ref{fig:e_th_independence} we demonstrate that the ground-truth dispersion obtained using this approach is nearly independent of the choice of $e_{\rm th}$. 

%%%%%%%%%%%%%%%%%%%%%%%%%%%%%%%%%%%%%%%%%%%%%%%%%%%%%%%%%%%%%%%%%%%%%%%
%%%%%%%%%%%%%%%%%%%%%%%%%%%%%%%%%%%%%%%%%%%%%%%%%%%%%%%%%%%%%%%%%%%%%%

%\bibliography{glass_refs.bib}

%apsrev4-2.bst 2019-01-14 (MD) hand-edited version of apsrev4-1.bst
%Control: key (0)
%Control: author (8) initials jnrlst
%Control: editor formatted (1) identically to author
%Control: production of article title (0) allowed
%Control: page (0) single
%Control: year (1) truncated
%Control: production of eprint (0) enabled
%

\end{document}